\documentstyle[aps,preprint,epsfig]{revtex}
\draft
\begin{document}
\bibliographystyle{prsty}
\makeatletter
\def\gsim{\compoundrel >\over\sim}
\def\lsim{\compoundrel <\over\sim}
\def\compoundrel#1\over#2{\mathpalette\compoundreL{{#1}\over{#2}}}
\def\compoundreL#1#2{\compoundREL#1#2}
\def\compoundREL#1#2\over#3{\mathrel
    {\vcenter{\hbox{$\m@th\buildrel{#1#2}\over{#1#3}$}}}}
\makeatother
\preprint{RUB-TPII-21/95}
\title{Elusive exotic states}
\author{ 
Hyun-Chul Kim
\footnote{{\it e-mail address}: kim@hadron.tp2.ruhr-uni-bochum.de}}
\address{Institut f\"ur Theoretische  Physik  II, \\  Postfach 102148,
Ruhr-Universit\" at Bochum, \\
 D-44780 Bochum, Germany  \\ }
\author{
Mikhail Shmatikov\footnote{{\it e-mail address}: msh@ofpnp.kiae.su}}
\address{ Russian Research Center "Kurchatov Institute", 123182 Moscow,
Russia}
\date{July 1995}
\maketitle
\begin{abstract}
The existence of flavor exotic $QQ\bar{q}\bar{q}$ molecular-type
states is investigated. An attractive force between
two {\it pseudoscalar} $H = (Q\bar{q})$ heavy meson is generated
by (correlated) two-pion exchange. The emergence of a (loosely)
bound state depends crucially on the value of the coupling
constant $g$ of the $H^*H\pi$ vertex. For a $g$ value calculated
from the experimental upper limit on the width of  the $D^*$ meson
the considered mechanism alone is strong enough to generate
a bound state in the $BB$ system while the $DD$ system is
very close to become bound. Such states, if exist, are
stable with respect to strong interactions. They may be
observed as stable scalar particles with the mass $M\approx 2\,m_H$ and
flavor quantum number $\pm 2$.
\end{abstract}
\vfill\eject
At present basically all the hadrons, both baryons and mesons, can be safely
classified as $qqq$ and $\bar qq$ quark states respectively. No reliable
evidence for multiquark states but nuclei, which can be considered as weakly
bound states of the baryons, has been found. One of the most clear-cut
signals for such multiquark states would be an observation of hadronic
states with exotic flavor quantum numbers. A promising hunting ground for
such an exotics is the domain of heavy flavors, or more precisely, hadrons
with the $QQ\bar q\bar q$ structure where $Q$ is a heavy quark. The sought
after hadronic states are known to exist in the $m_Q\rightarrow \infty $
limit \cite{Richard,Lipkin,Manohar}. Quantitatively the heavy-mass limit can
be recast in the form of an inequality 
\begin{equation}
\alpha _s^2(m_Q)\,m_Q\ \compoundrel >\over\sim \Lambda _{QCD}\,.
\end{equation}
However, this inequality is satisfied for the $t$-quark only which, on the
other hand, decays before the hadronization. For $Q=c,b$ quarks $QQ\bar
q\bar q$ hadrons may exist as weakly-bound systems of two $Q\bar q$ mesons.

A couple of $Q\bar q$ mesons may be bound, for the large enough mass of a
heavy quark, by a comparatively weak force generated by one-pion
exchange. The attractive feature of such a long-range potential is that it is
calculable in a chiral perturbation theory. A molecule-type $H$ -- $H$ ($H$
= $Q\bar q$) hadron exists in the limit of the infinitely
heavy mass $m_Q$ \cite{Manohar}. Corrections $\sim 1/m_Q$, in the realistic
case of $b$- or $c$- quarks, prove to be of importance. 
The investigation of such systems in \cite{Manohar,T1,T2} 
yielded rather controversial results.  
According to \cite{Manohar} the $D$ -mesons are too light to be bound by the
one-pion-exchange force, whereas a loosely bound state was found in the
system of two $B$-mesons consisting of $BB^{*}$ and $B^{*}B^{*}$ with equal
weights. At the same time the one-pion-exchange force between two
heavy mesons was found in \cite{T1,T2} to be too weak (or even repulsive) to
produce a bound state. In contrast, systems with {\it non-exotic} quantum
numbers were shown to have a rich spectrum of bound states: $D\bar D$ system
is expected to have a bound state and the $B\bar B$ one possesses several such
states with various spin-parity quantum numbers. It should be stressed that
conclusions as to the existence of a bound state(s) rely heavily on the
specific value of the strength constant $g$ of the $\pi $-meson coupling to
a heavy meson which sets the overall scale of the interaction strength (see
below).

In the limit $m_Q \rightarrow\infty$ mesons containing a single heavy quark
come in degenerate doublets of pseudoscalar ($H$) and vector ($H^*$) mesons.
Mass corrections being taken into account, 
the degeneracy in the total momentum is lifted off and the $H$ and 
$H^*$ mesons are to be treated separately. For
this reason one-pion exchange is operative in the $H^*H^*$ and $HH^*$ (or $%
\bar{H}^*H^*$ and $\bar{H}H^*$) systems only and all the conclusions of \cite
{Manohar,T1,T2} refer just to such systems.

In the present paper we investigate coupling and possible bound states of
two {\it pseudoscalar} heavy mesons $HH$. There are two reasons arousing
interest to such a system. First, since a pseudoscalar $H$ -meson is lighter
than its vector counterpart $H^{*}$, a bound $H$--$H$ state will be stable
with respect to a strong decay. An $H^{*}$ -- $H^{*}$ system, spin-parity
allowing, may decay into the $HH$ pair. Open decay channel brings additional
repulsion which may destroy a bound state \cite{SH}. Another reason for
investigating a system of two pseudoscalar meson has a dynamical nature. 
The vertex of pion coupling to a pair of the $H$-mesons apparently vanishing,
the next-in-range force operative in this system is (correlated)
two-pion exchange (CTPX). This force is known to provide the bulk of the
attraction in the $NN$ system \cite{Ericson}.  Thus it is of interest to
investigate its effects in heavy-meson systems, and the $HH$ pair provides a
testing ground where the CTPX forces are not obscured by the one-pion
exchange.

The construction of the CTPX potential is described in the plethora of works
(see \cite{Ericson,Rho,Brown} just to mention a few). The generic form of 
two-pion exchange is depicted in fig.1. 
A pair of pions in the $t$-channel
may have various orbital momenta and isospin. We focus our attention on the 
$J^\pi (T)=0^{+}(0)$ channel, which is the most relevant for the system under
consideration, since it is responsible for the strong sttraction.
We will postpone the discussion of $J\geq 1$ contributions to the
conclusion. As is well known from the works on the $NN$ interaction, 
the static potential for the scalar-isoscalar channel can be written by
\begin{equation}
V_{2\pi }(r)=-\frac 1\pi \int_{4m_\pi ^2}^{t_{max}}\,dt\rho (t)\,
\frac{e^{-\sqrt{t}r}}{4\pi r}.                         
\label{pot}
\end{equation}
\vskip18pt
\centerline{\epsfysize=1.8in\epsffile{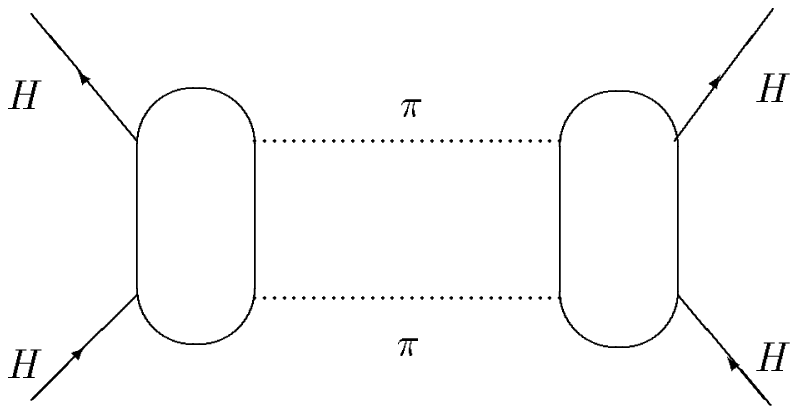}}\vskip4pt
{\bf Fig.1}: Two-pion intermediate state in the $HH\rightarrow HH$ scattering.
\vskip6pt
The dynamics of the intercation is controlled by the behavior of the spectral
function $\rho (t)$. It vanishes at the two-pion treshold 
in the $t$-channel, i.e. at $t=4m_\pi ^2$. The upper integration
limit $t_{max}$ is usually chosen $t_{max}\simeq 50 m^2_{\pi}$. 
Thus the CTPX potential is a
superposition of Yukawa forces corresponding to the exchange by a meson with
the $\sqrt{t}$ mass and the weight of this configuration is given by the
corresponding value of the $\rho (t)$ spectral function.

The spectral function is in turn expressed, by means of the unitarity
condition, in terms of the amplitude ${\cal A}$ of the $\bar HH\rightarrow
\pi\pi $ process: 
\begin{equation}
\rho (t)=\frac 1{32\pi }\sqrt{\frac{t-4m_\pi ^2}t}|{\cal A}(t)|^2.
\label{spect}
\end{equation}
Note that in contrast to the entirely familiar $\bar NN\rightarrow \pi \pi $
case there is no iterated Born term to be subtracted. 
In case of the $\bar NN\rightarrow \pi \pi $ process, the quasiempirical 
data in the pseudophysical region ($4m^2_{\pi}< t < 50m^2_{\pi}$) can
be obtained by making an analytic continuation of the $\pi N$ 
scattering data.  However, for lack of the quasiempirical information 
on the $\pi$-meson scattering off a heavy meson we are forced 
to apply a dynamical model.     

Such a model developed in \cite{Kim} for the $NN$ interaction was shown to
reproduce with good accuracy available experimental data on scattering phase
shifts in a wide energy region.  The dynamical model tailored for the case of 
$\bar HH\rightarrow \pi \pi $ amplitude is shown in fig.2. 
\vskip18pt
\centerline{\epsfysize=1.8in\epsffile{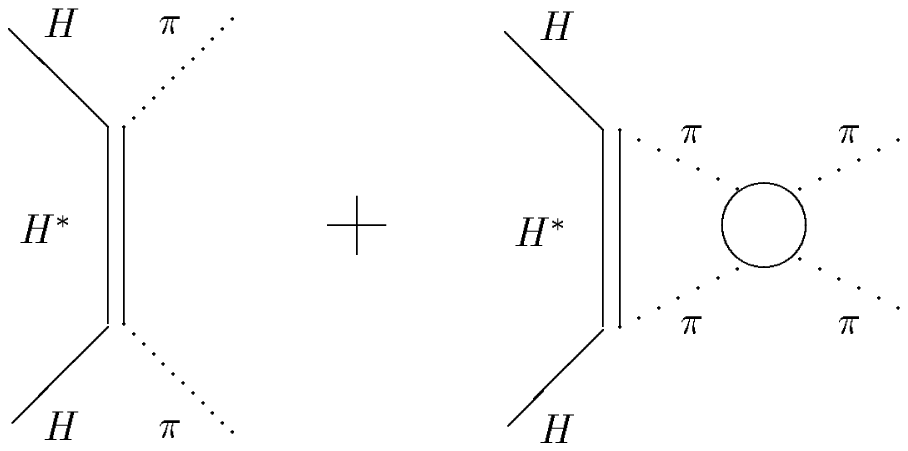}}\vskip4pt
{\bf Fig.2}: Dynamical model of the $\bar{H}%
H\rightarrow \pi\pi$ annihilation amplitude. Empty circle denotes the
T-matrix of the $\pi\pi\rightarrow \pi\pi$ rescattering.
\vskip10pt 
 It consists of two components: 
the Born term where outgoing pions do not interact with each
other and the one involving pion rescattering. The amplitude of the
(half-off-shell) $\pi \pi \rightarrow \pi \pi $ interaction can be
evaluated by using the meson-exchange model of 
$\pi \pi $, $\pi K$ and $KK$ processes~\cite{Lohse}. 
 As usual, a nonvanishing range of the strong interaction is
taken into consideration by introducing a formfactor $F$ in the $HH^{*}\pi $
vertex.  We choose the exponential form \cite{Mous} for it:
\begin{equation}
F(k^2)=\exp \left\{ -\frac{k^2+M^2}{2\Lambda }\right\}  \label{ff}
\end{equation}
where $M$ is the mass of the exchange particle 
and $\Lambda $ denotes the cut-off parameter.

The overall strength scale of the (\ref{pot}) potential is established by
the constant $g$ of the $\pi H^{*}H$ coupling. It is defined as follows \cite
{Wise} 
\begin{equation}
{\cal L}=\frac i2\,g\,{\rm Tr}\left[ H\gamma _\mu \gamma _5\left( \xi
^{\dagger }\partial _\mu \xi -\xi \partial _\mu \xi ^{\dagger }\right) \bar
H\right]
\end{equation}
In principle, the value of $g$ could be determined from the width of the $%
H^{*}\rightarrow H\pi $ decay channel. However, because of small mass
difference $\Delta m_B$ between $B$ and $B^{*}$ mesons ($\Delta m_B\approx
46 $ MeV) this channel is closed for the $B^{*}$-meson. As to the system of
(vector and pseudoscalar) $D$-mesons, at present the upper limit for the
total width of the $D^{*}$-meson and partial width of the $D^{*}\rightarrow
D\pi $ decay are measured \cite{exp}. Theoretical estimates of the $g$ value
are controversial. The upper limit in both the chiral and the heavy quark
limits in the leading $1/N_c$ order reads \cite{Maxim}: 
\begin{equation}
g\leq \frac 1{\sqrt{2}}\left[ 1+{\cal O}(1/m_H)+{\cal O}(1/N_c)+{\cal O}%
(m_\pi m_H)\right]
\end{equation}
The numerical values utilized in \cite{T1,T2} and \cite{Manohar}, which
ensured the existence of a (loosely) bound state, are equal to $g\approx 0.6$
and $0.7$ respectively. These values were obtained by assuming that the upper
experimental limit on the width of the $D^{*}$ meson is just the realistic
value of the latter. Direct calculations of $g$ based on the QCD sum
rules \cite{Pietro} yielded much smaller value: $g\approx 0.2$. Note that
the $H^{*}H\pi $ vertex enters in the spectral function 
$\rho (t)$ four times, which
implies that the CTPX potential $V_{2\pi }$ (\ref{pot}) contains the $g^4$
factor. Thus the specific value of $g$ proves to be crucial for the
(non)existence of a bound state of two pseudoscalar $H$ -- $H$ mesons.

Under these circumstances we have investigated the existence of the $H$ -- $%
H $ bound state and, if present, calculated the binding energy $E_b$ in a
wide range of $g$'s: $0\leq g\leq 1$. In the numerical calculations of the
CTPX potential (\ref{pot}) the value of $t_{max}$ was chosen, as it is
customary in the $NN$ system, equal to $t_{max}=50\,m_\pi ^2$. The spectral
function (\ref{spect}) being positive-defined, a decrease of the $t_{max}$
may be compensated by the corresponding increase of the $g$ coupling
constant. The cut-off parameter $\Lambda $, entering into the formfactor (%
\ref{ff}), was varied in the $\Lambda =1.5\div 2.1$~GeV interval.

\vskip18pt
\centerline{\epsfysize=2.5in\epsffile{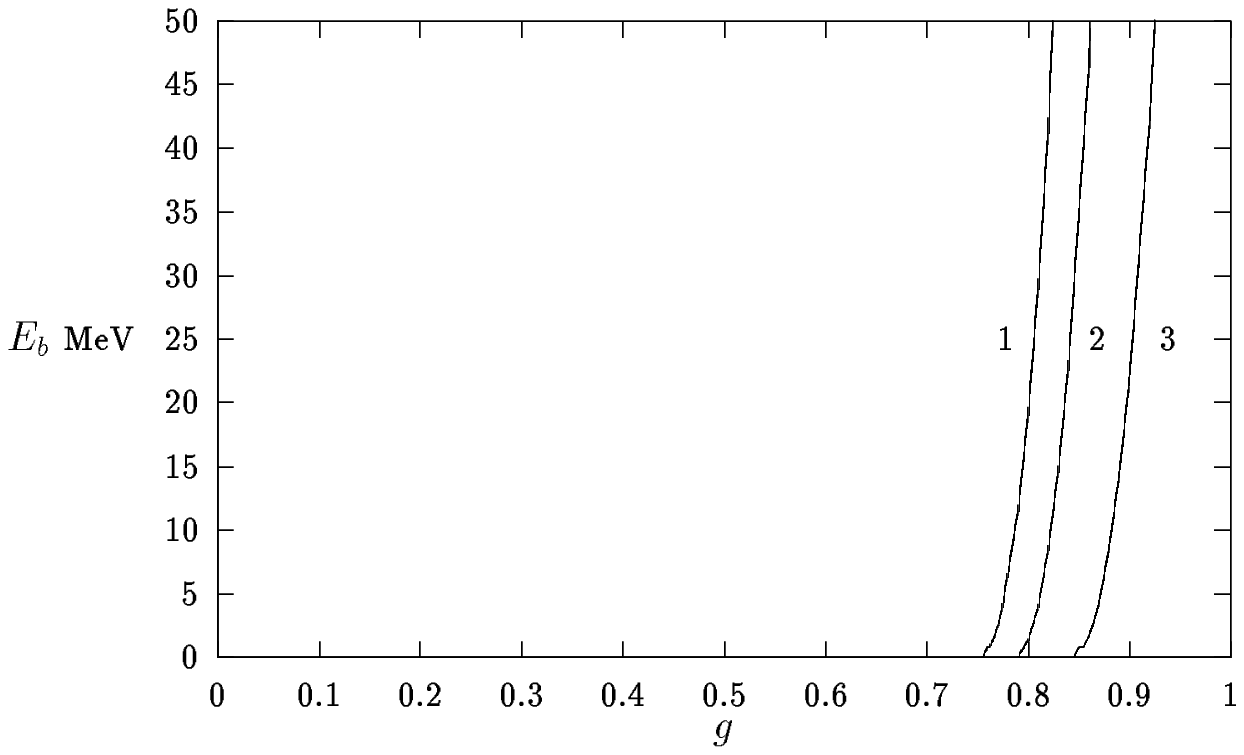}}\vskip4pt
{\bf Fig.3}: Dependence
of the binding energy $E_b$ of two pseudoscalar $D$-mesons on the $D^*D\pi$
coupling constant $g$. Curves labeled 1,2 and 3 correspond to the cut-off
paremeter $\Lambda = 2.1$, 1.8 and 1.5 GeV respectively.
\vskip6pt
Results of calculations for the $DD$ system ($S$-wave) are presented in
fig.3. Note the (anticipated) sharp dependence of the binding energy on the $%
g$ value and its rather weak spread with the variation of $\Lambda $. A
conclusion, which can be drawn from the investigation of the $g$ dependence,
is that the bound state emerges at the value of $g\approx 0.75\div 0.85$,
i.e. at somewhat larger $g$'s than the molecular states of refs. \cite
{T1,T2,Manohar}. In the case of more heavy $B$-mesons situation proves to be
more favorable (see fig.4). Here the bound state emerges, depending upon the
cut-off parameter, in the $g\approx 0.58\div 0.78$ range.

\vskip18pt
\centerline{\epsfysize=2.5in\epsffile{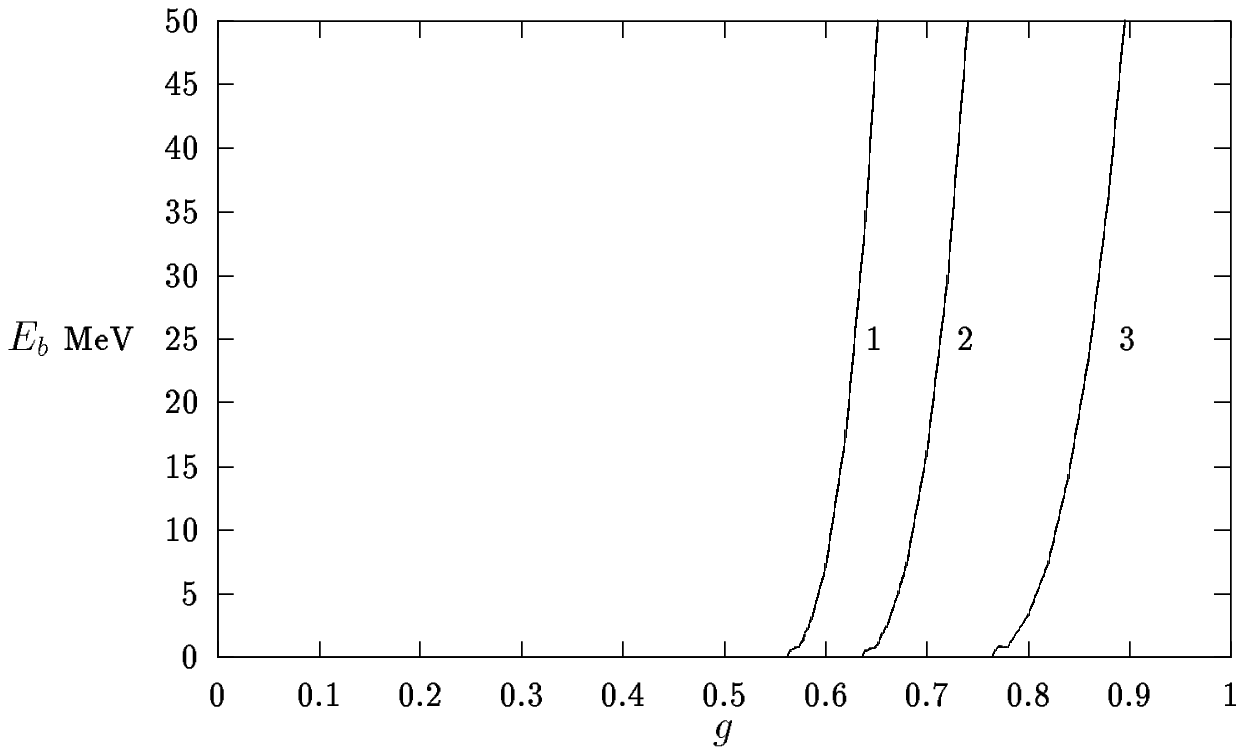}}\vskip4pt
{\bf Fig.4}:  Dependence
of the binding energy $E_b$ of two pseudoscalar $B$-mesons on the $B^*B\pi$
coupling constant $g$. Curves labeled 1,2 and 3 correspond to the cut-off
paremeter $\Lambda = 2.1$, 1.8 and 1.5 GeV respectively
\vskip6pt
Hence, one can conclude that flavor exotic $DD$ (molecular) states are rather
non-existing. Two pseudoscalar $B$ mesons become bound at about the same
value of the $g$ coupling constant when the bound state in the $\bar
B^{*}B^{*}$ and $\bar B^{*}B$ \cite{T1,T2} or $B^{*}B^{*}$ \cite{Manohar}
systems emerges. For small $g\approx 0.2$ \cite{Pietro} molecular-type
states with any meson content do not exist. It should be noted, that in
contrast to the case of one-pion exchange 
which has the opposite (because of
G-parity) sign in the $H^{*}H^{*}$ and $\bar H^{*}H^{*}$ systems, CTPX
remains attractive in both meson-meson and meson-antimeson systems. It
implies that obtained results hold true also for the $\bar DD$ and $\bar BB$
(non-exotic) systems. Stated differently, provided the latter molecular
states are observed, existence of their {\it flavor exotic} counterparts ($%
BB $ and $DD$) is highly plausible. Considered mechanism is operative as
well in similar system containing vector heavy mesons. The main difference
is that one-pion exchange is operative in such systems. Then the
attraction provided by the CTPX can either in combination with the
one-pion-exchange induced attraction or overwhelming repulsion, generate
otherwise unobtainable molecular-type states.

We have considered the possible formation of molecular-type bound 
states of two pseudoscalar heavy mesons generated 
by the CTPX with the $t$-channel quantum
numbers $J^\pi =0^{+}$. The effective potential $V_{2\pi }$ 
is contributed as well by the CTPX with other quantum numbers 
$J^\pi =1^{-}$ and $2^{+}$.
However, the $\pi \pi $ interaction with such quantum numbers involves
additional, as compared to the considered case, momenta of pions. The
integration over the pionic loop (fig.2) will result then in a spectral
function peaked at larger values of $t$ or, in the coordinate space, to the
CTPX potential with smaller interaction range. Barring accidental emergence
of a state with almost vanishing binding energy such states are expected to
be compact ones. Their dynamics will be governed by the combined action of
CTPX and QCD quark-quark interaction. Existence of bound states in the
situation when both meson exchanges and quark coupling come into play will
be considered elsewhere.

Summarizing we have investigated a possible existence of molecular-type bound
states in flavor exotic $HH$ systems. In a sense it is an extension of the
results obtained in \cite{Manohar} which predicted, in the $m_Q\rightarrow
\infty $ limit, the existence of a bound heavy-meson state. However, $1/m_Q$
corrections are of importance, necessitating the different approach to the
vector $H^{*}$ and pseudoscalar $H$ heavy mesons. All the heavy-meson
systems differ drastically from the one ($HH$) considered in the present
paper. First, their dynamics is controlled by the combined action of the
one- and (correlated)\ two-pion exchange. It implies that besides the $%
H^{*}H\pi $ coupling constant one more strength coupling constant (in the $%
H^{*}H^{*}\pi $ vertex) plays an important role.  Second, many of the states
involving the vector $H^{*}$ heavy meson may decay due to the $%
H^{*}\rightarrow H$ transformation. The presence of an open decay channel
may affect strongly the properties and the very existence of the sought
after bound state \cite{SH}. At the same time coupling of the considered
$HH$ system to (closed) channels containing vector heavy mesons makes
the binding of two pseudoscalar mesons more strong. Thus, the search of a
possible molecular-type
state in any heavy-meson system but the one ($HH$) considered in the present
paper requires much more extensive analysis, making possible conclusions
substantially less unambiguous.

The existence of a flavor-exotic $HH$ tetraquark depends crucially on the
value $g$ of the $H^{*}H\pi $ coupling constant. The latter is related to
the width of the $H^{*}\rightarrow H+\pi $ decay and, since corresponding
partial width is measured, on the total width $\Gamma $ of the $H^{*}$%
-meson. Provided $\Gamma $ is taken equal to the experimental upper limit,
the CTPX induced forces alone are strong enough to produce a bound state of $%
BB$ mesons, while the $DD$ system is very close to support a bound state.
Such bound states will manifest themselves as stable scalar mesons with the
mass $M\approx 2\,m_H$ and the flavor quantum number equal to $\pm 2$ (note
that existence of a $HH$ bound state implies that the $\bar H\bar H$ system
is also bound). The same conclusion holds true for the flavor-hidden $\bar
hH $ states which have the same binding energy. It is relevant to note that
CTPX induced {\it flavor exotic} molecular-type states emerge at about the
same values of $g$ which ensure boundedness of {\it flavor-hidden} $\bar
H^{*}H^{*}$ and $\bar H^{*}H$ states generated by one-pion exchange \cite
{T1,T2,Manohar}. At smaller values of $g$ molecular-type states with any
heavy meson content are not expected to emerge. More complicated systems of
heavy mesons where both one- and (correlated) two-pion exchanges are
operative will be considered elsewhere.

\end{document}